\documentclass[pra,aps,showpacs,groupedaddress,superscriptaddress,twocolumn,toc=flat,biblatex,footinbib]{revtex4-1}
\usepackage{natbib}
\usepackage[table]{xcolor}
\usepackage[colorlinks=false]{hyperref}
\usepackage[utf8]{inputenc}
\usepackage{color}
\usepackage{bbm} 
\usepackage{bbold}
\usepackage[english]{babel}
\usepackage{multirow}
\usepackage{amsfonts,amsmath,amssymb,stmaryrd}
\usepackage{placeins}
\usepackage{graphicx}
\usepackage{subfigure}  
\usepackage{braket}
\usepackage{epsfig}
\usepackage{mathrsfs}
\usepackage{verbatim}
\usepackage{centernot}
\usepackage{ulem}
\usepackage{longtable}
\usepackage{nicefrac}
\usepackage[framemethod=tikz]{mdframed}
\usepackage{siunitx}
\usepackage[nolist]{acronym}

\usepackage{soul}
\usepackage{array}

\usepackage{cancel,ifthen}
\newcommand{\cmnt}[2][NoInPuT]{\ifthenelse{\equal{#1}{NoInPuT}}{}{{\color{red}\sout{#1}}} {\color{blue} #2}}

\usepackage{bm}	
\renewcommand{\vec}[1]{\bm{#1}}

\LTcapwidth=\textwidth

\begin{document}

\title{
Atomistic spin dynamics simulations of magnonic spin Seebeck and spin Nernst effects in altermagnets
}

\author{Markus Weißenhofer}
\email[]{markus.weissenhofer@fu-berlin.de}
 \affiliation{Department of Physics and Astronomy, Uppsala University, P. O. Box 516, S-751 20 Uppsala, Sweden}
\affiliation{Department of Physics, Freie Universit{\"a}t Berlin, Arnimallee 14, D-14195 Berlin, Germany}

\author{Alberto Marmodoro}
\affiliation{Institute of Physics, Czech Academy of Sciences, Cukrovarnická 10, 162 00 Praha 6, Czech Republic}


\date{\today}

\begin{abstract}
Magnon band structures in altermagnets are characterized by an energy splitting of modes with opposite chirality, even in the absence of applied external fields and relativistic effects, due to an anisotropy in the Heisenberg exchange interactions. We perform quantitative atomistic spin dynamics simulations based on ab initio electronic structure calculations on rutile RuO$_2$, a prototypical "d-wave" altermagnet, to study magnon currents generated by thermal gradients. We report substantial spin Seebeck and spin Nernst effects, i.e., longitudinal or transverse spin currents, depending on the propagation direction of the magnons with respect to the crystal, together with a finite spin accumulation associated with non-linearities in the temperature profile. Our findings are consistent with the altermagnetic spin-group symmetry, as well as predictions from
linear spin wave theory and semiclassical Boltzmann transport theory.
\end{abstract}
\maketitle

\section{Introduction}
The study of magnons, low energy collective excitations of magnetic systems, offers important insight into fundamental properties of condensed matter systems. In addition to that, magnons have been extensively explored in light of their potential for novel applications  \cite{Kajiwara2010,Bauer2012,Chumak2014,Chumak2015,Cornelissen2015,Pirro2021}.
These efforts lead to the foundation of the field of \textit{magnon spintronics}, which aims at developing energy-efficient information storage and processing strategies by eliminating energy loss due to Joule heating associated with charge transport.

Magnons in ferromagnetic materials are characterized by a well-defined chirality, as a result of time reversal symmetry breaking, and can hence carry spin currents, enabling their usage for magnon spintronics \cite{Kajiwara2010}. However, their dispersion relation is typically quadratic, rendering the group velocity wave-vector-dependent and thus hindering the propagation of stable magnon wave packets. 

Antiferromagnetic magnons, on the other hand, typically have a linear dispersion relation close to the center of the Brillouin zone, as long as relativistic effects are neglected, and reach the THz regime \cite{Gomonay2014,Gomonay2018}. Due to conserved symmetry under inversion and time reversal, the corresponding two magnon modes with left- and right-handed chirality are degenerate across the entire Brillouin zone in conventional collinear antiferromagnets \cite{Rezende2019}. Hence, observation of magnon currents carrying finite spin angular momentum requires either an externally applied magnetic field \cite{Rezende2016,Li2019,Rezende2019b}, or the presence of higher order exchange interactions, such as e.g.~ the Dzyaloshinsky-Moriya interaction. The latter introduces a directional modulation (or non-reciprocity) of the dispersion relation
\cite{Udvardi2009,Zakeri2010}. The magnitude of this modulation will however depend, approximately, on the strength of spin-orbit coupling (SOC).

\begin{figure}
    \centering
        \includegraphics[trim=2.5cm 0.0 1.0cm 1.0cm, clip,width=1.0\linewidth]{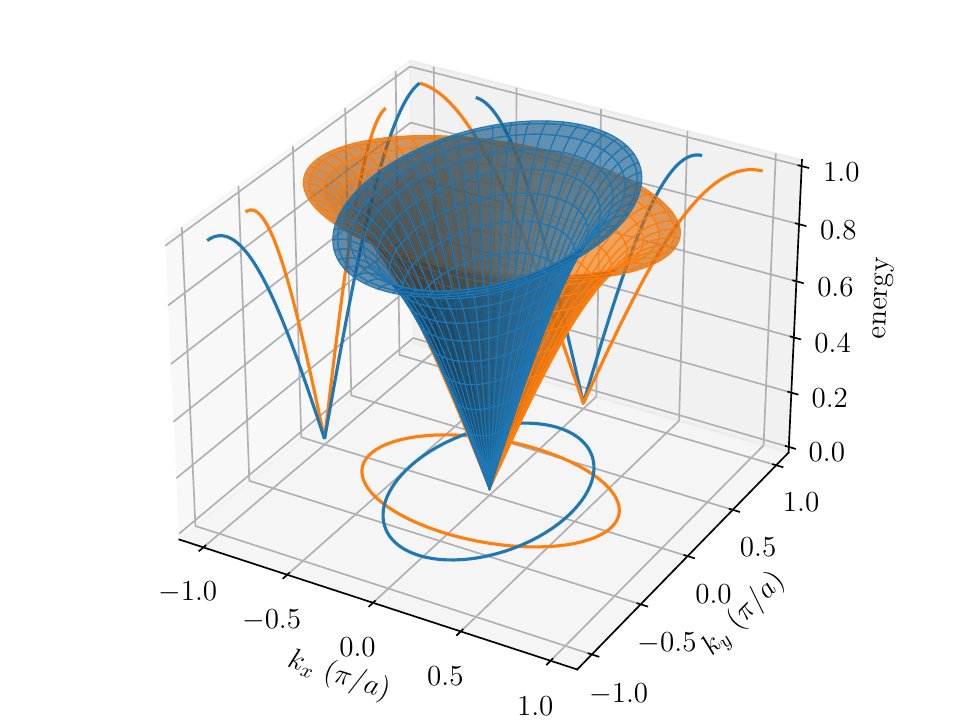}
    \caption{Schematic representation of the magnon dispersion in "d-wave" altermagnets,
    such as in particular in the case of rutile unit cells. 
    The dispersion of both magnon modes is linear around the $\Gamma$ point, with an alternating frequency split across the Brillouin zone. The solid lines correspond to sections through the dispersion relation for some constant energy (projected onto the $k_x,k_y$ plane), for $k_y=0$ ($k_x$, energy plane) and for $k_x=0$ ($k_y$, energy plane).
    }
    \label{fig:sketch_dispersion}
\end{figure}

We turn therefore our attention to altermagnetism, which has been recently established as a special case of collinear antiferromagnetism that is characterized by a spin split band structure for electrons \cite{Smejkal2022,Smejkal_2022b} and magnons \cite{Smejkal2023_chiral_magnons,Cui2023}, even in the absence of relativistic effects and external fields. In the presence of crystal-field splitting acting along different directions on antiparallel magnetic atoms, the (scalar) Heisenberg exchange interactions can also pick up a sizable direction-dependence i.e. anisotropy, even regardless of SOC \cite{Hayami2022}. As a consequence, one can predict that the magnon dispersion will be in general non-degenerate across the Brillouin zone, except for marginal set of $\vec k$ points \cite{Herring1937,Smejkal2022} (corresponding to e.g. nodal planes, and depending on the specific spin-group symmetry of the ground state), as schematically illustrated in Fig.~\ref{fig:sketch_dispersion}.

As for any collinear antiferromagnet, the wave-like precession of atomic magnetic moments
will involve to variable extent, in general, the different magnetic sublattices 
as a function of $\vec k$: approaching the boundary of the Brillouin zone,
magnons will tend to be hosted exclusively by one or by the other sublattice,
while they will equally involve both of them 
toward the $\Gamma$ point.
Away from it, the different 
and now energy-split two magnon dispersions
describe excitations that carry (spin) angular momentum
with opposite sign \cite{Sahni1974,Smejkal2023_chiral_magnons,McClarty2023,Maier2023}.
This can pave way to a variety of concept applications
revisiting earlier proposal initially put forward
through reliance only on the different, 
higher-order Heisenberg exchange interactions route
in order to achieve non-reciprocity \cite{Jamali2013}. 

Of course, in a real sample both SOC-driven and crystal-field-driven mechanisms for anisotropic exchange interactions would normally occur concurrently.

In this work, we aim to investigate magnon currents in altermagnets by means of quantitative ab initio electronic structure and
atomistic spin dynamics calculations, as well as analytical
linear spin-wave theory derivations.
We will study the steady spin current 
set in motion by a thermal gradient between two heat reservoirs
and mediated by magnons, i.e. the magnon spin Seebeck 
and spin Nernst effects \cite{Cui2023}.

In particular, we will examine how these emerge
as a function of relative direction 
between temperature gradient,
anisotropy in the (scalar) Heisenberg exchange interactions,
and the resulting non-degenerate magnon dispersion.
This scenario is realized by altermagnets
in full analogy with their energy split and alternating
spin-polarized electronic band structure 
and beside possible further effects proportional to SOC \cite{Hayami2020,Yuan2021},
due to the interplay between orientation of crystal field splitting from non-magnetic atoms
which introduces strong anisotropy in the magnetization density
\cite{Pekar1964,Yuan2021a,Smejkal_2022a},
and the direction of anti-parallel magnetic sublattices.

We adopt as prototype material for our 
calculations 
the example of rutile RuO$_{2}$. 
While other studies have started to examine
electronic-mediated thermal transport \cite{Zhou2023},
we focus here on the magnonic degree of freedom of altermagnets. In addition to its thermally-generated, magnon-mediated steady-state spin currents
between two heat reservoirs,
our simulations also predict the emergence of a finite local magnetic moment, or spin accumulation, 
in correspondence to non-linearities in the temperature gradient profile.

This paper is structured as follows.
We begin by recalling the essential features of this material, 
highlighting those aspects that qualify it as an altermagnet.
In particular, details 
for the spin model parameters 
and the ensuing magnon dispersion for RuO$_{2}$ 
before applying out-of-equilibrium thermal gradient
are given in Sec.~\ref{sec:RuO2}.

Theoretical methods used for the simulation of magnon dynamics due to temperature gradients
in terms of a stochastic Landau-Lifshitz-Gilbert equation of motion
are briefly introduced in Appendix~\ref{sec:app_methods}.

We then report numerical results 
from such
atomistic spin dynamics simulations (Sec. ~\ref{sec:results}).
These are also confirmed 
through comparison with predictions 
from linear spin wave theory, 
in combination with semiclassical Boltzmann transport theory (Appendix~\ref{sec:app_analytic}). 

We then summarize our findings and offer our conclusions (Sec.~\ref{sec: conclusions}).

\section{Magnon dispersion in RuO$_{2}$}
\label{sec:RuO2}
We perform numerical calculations
based on ab initio electronic structure 
and atomistic spin dynamics,
in order to provide a first realistic estimate
of magnon-mediated Seebeck and Nernst effect 
associated with the non-relativistic component of non-degenerate spin-wave dispersion 
in altermagnets \cite{Smejkal2023_chiral_magnons,McClarty2023}.

We choose in particular RuO$_2$ as prototype altermagnet \cite{Fedchenko2023},
due to the growing body of theoretical and experimental literature dedicated to this compound.
This material received initial attention
due to metallic conductivity,
relatively rare for oxides, 
in combination with reports
of collinear antiferromagnetic order up to room temperature \cite{Berlijn2017},
leading to high expectations for possible
electron-mediated spintronics applications \cite{Yuan2020,Bose2022,Bai2022,Karube2022,Zhou2023}.

On the other hand, this work focuses on transport properties mediated by magnons, not by electronic charge-carriers. Other spin-split antiferromagnets
(or altermagnets) with a gap at the Fermi level,
but with alike unit cell geometry and magnetic order
(such as e.g. MnF$_2$, FeF$_2$ or CoF$_2$, 
just to remain within the subclass of rutile structures),
would also be valid candidate materials.
Our conclusions will also apply, qualitatively, 
to cases with more than two sublattices with antiparallel magnetic moments
and/or with a different than four-fold rotation symmetry, 
such as for the example of ferric oxide Fe$_3$O$_4$, 
or of MnTe \cite{Smejkal2023_chiral_magnons}.

\begin{figure}[ht!]
    \centering
    \includegraphics[width=1.0\linewidth]{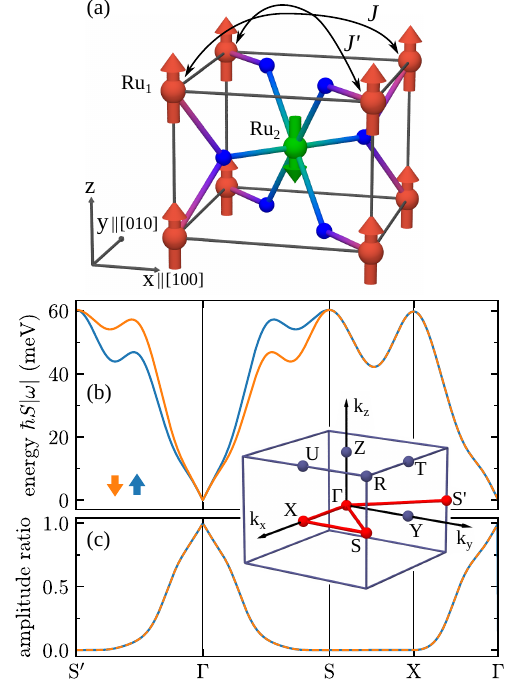}
    \caption{(a) Unit cell of the altermagnetic rutile phase of RuO$_2$.
    Collinear antiparallel magnetic order across Ru sublattices 
    is represented via red and green colors and arrows. Non-magnetic 
    O atoms are depicted as smaller blue spheres. 
    The most significant anisotropic Heisenberg exchange interactions $J,J'$ are also highlighted via thin black arrows. 
    (b) Magnonic band structure of RuO$_2$ shown along a Brillouin zone path (inset, highlighted in red) within the $k_x-k_y$ plane crossing the $\Gamma$ point. 
    The bold arrows indicate the net spin of the modes. 
    (c) Corresponding amplitude ratio between the tilting of the Ru moments at each sublattice.}
    \label{fig:dispersion}
\end{figure}

RuO$_{2}$ crystallizes in the rutile structure, see Fig.~\ref{fig:dispersion}a, with two opposite-spin sublattices~\cite{Berlijn2017}, as further discussed also in Refs.~\cite{Zhu2019,Lovesey2022,Lovesey2023}. Due to the non-magnetic atoms
occupying the Wyckoff position 4f,
equivalence between Ru$_1$ and Ru$_2$ cannot be achieved 
through pure time reversal, i.e. flipping the sign of the N\'{e}el vector,
but only when this symmetry operation 
is also accompanied by a space rotation,
by 90$^{\circ}$ in this rutile example geometry.
The requirement of such combination of time and space symmetry operations
characterizes altermagnets, in comparison with conventional collinear antiferromagnets.

In this work we adopt
RuO$_{2}$ 
Heisenberg exchange parameters
calculated from density functional theory
by means of the magnetic force theorem,
in similar fashion as Ref.~\cite{Smejkal2023_chiral_magnons}.

In the corresponding Heisenberg Hamiltonian
\begin{align}
\label{eq:Hamiltonian}
\mathcal{H}
=
-
\sum_{ij}
\sum_{ss'}
J_{ij}^{ss'}
\vec{S}_{is}
\cdot
\vec{S}_{js'}
,
\end{align}
$J^{ss'}_{ij}$ denotes scalar coupling constants, and $\vec{S}_{is} = \vec \mu_{is} / | \mu_s$ is a unit vector describing the direction of the magnetic moment $\vec \mu_{is}$ for the magnetic atoms at position $\vec{r}_i+\vec{b}_s$, with $\vec{r}_i$ being the location of the unit cell within the periodic lattice, and $\vec{b}_s$ the sublattice position within the unit cell. 

Notably, exchange constants between atomic magnetic moments at relative position $\vec{R}^{ss'}_{ij}=\vec{r}_j-\vec{r}_i+\vec{b}_{s'}-\vec{b}_s$ not only depend on their distance $|\vec{R}^{ss'}_{ij}|$,
but also on the direction of $\vec{R}^{ss'}_{ij}$. 
In this particular example, this is mainly due
to the occurrence of O atoms
in the low-symmetry Wyckoff 4f position. 
Their biggest influence can be noted for the intra-sublattice exchange constants 
between Ru atoms at a distance of $|\vec{R}^{ss'}_{ij}|=\sqrt{2}A_\mathrm{lat}$ (with $A_\mathrm{lat}$ length of the edge for the square unit cell section in the $x,y$ plane of a rutile lattice). 
This displacement vector either intersects
oxygen atoms along the $[110]$ direction,
or it passes in-between them along the orthogonal
$[\bar{1}10]$ direction (Fig.~\ref{fig:dispersion}a).

\begin{figure*}[ht!]
    \centering
    \includegraphics[width=1.0\linewidth]{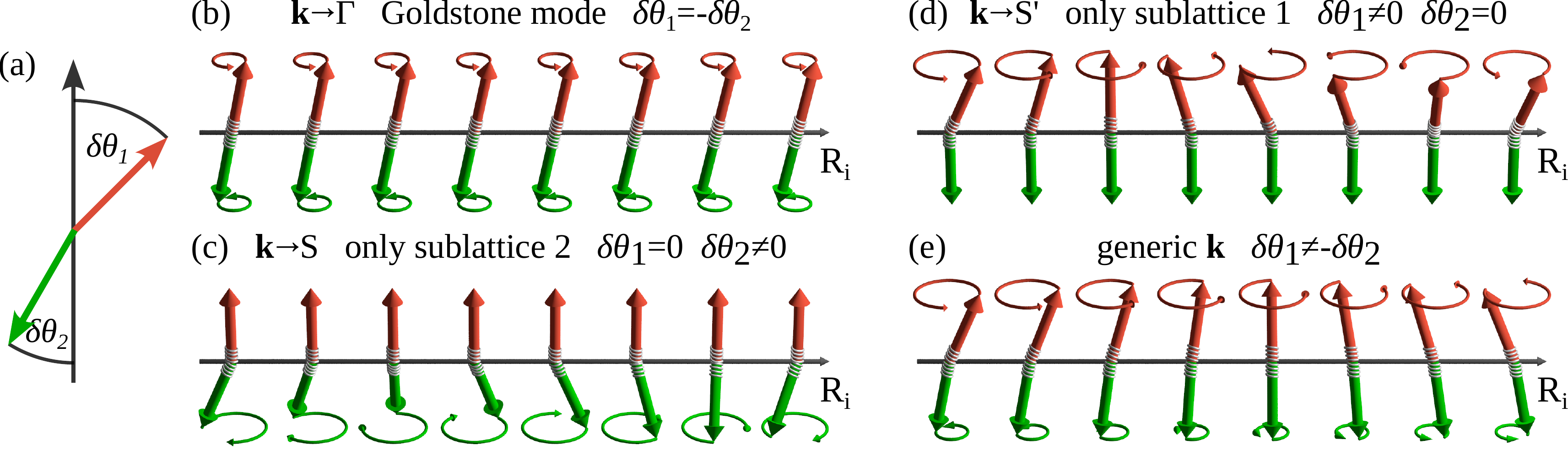}
    \caption{Schematic depiction of the precessing atomic magnetic moments from  antiparallel Ru sublattices (red or green arrows), belonging to a magnon eigenmode with wave-vector $\vec k$. 
    (a) Their deviation from the ground-state magnetic order is given by the azimuth angle $\delta \theta$ via $S^z_{n}=\cos (\theta_{n} + \delta\theta_{n})$, where $\theta_1=0$ and $\theta_2=\pi$ for the two magnetic sublattices.
    (b) In the $\vec k \to \Gamma$ limit, both tilts are the same.
    (c, d) Upon approaching the Brillouin zone boundary,
    the excitations involve predominantly only one or only the other sublattice,
    with largest difference along $\Gamma \to S$ or vice-versa along the orthogonal $\Gamma \to S'$ directions.
    (e) For generic $\vec k$, both tilt angles are finite but different.
    }
    \label{fig:eigenmodes limits for rutile altermagnets}
\end{figure*}

We calculate here the magnon band structure 
from the spin Hamiltonian Eq.~\eqref{eq:Hamiltonian}, using the Holstein-Primakoff transformation up to second order in magnon variables
and diagonalizing the resulting Hamiltonian 
via the Bogoliubov–Valatin transformation \cite{Rezende2019}. 
The eigenenergies of both magnon branches along different paths 
of the $k_x,k_y$ Brillouin zone section  
are shown in Fig.~\ref{fig:dispersion}b for $k_z=0$. 
Similarly to the results of Ref.~\cite{Smejkal2023_chiral_magnons},  
we obtain a linear dispersion close to the $\Gamma$ point, 
and a characteristic energy splitting of the magnon branches 
away from nodal manifolds of the Brillouin zone, 
which for this rutile geometry coincide 
for instance with the $k_x,k_z$ or with the $k_y,k_z$ planes. 
The corresponding eigenfrequencies have opposite sign 
along each of the two branches, 
i.e., the Ru moments of the first sublattice rotate in opposite direction with respect to the moments of the second sublattice (Fig.~\ref{fig:eigenmodes limits for rutile altermagnets}).

This angular velocity defines the chirality of each branch,
and it corresponds to the angular momentum carried by each magnon mode.
We denote here as right-handed the counterclockwise rotation of the Ru moment on the first sublattice,
and conversely we denote as left-handed the opposite, clockwise rotation. 

These two possible chiralities are accompanied by a different tilting 
away from the ground state direction
for the Ru moments at each sublattice. 
Upon approaching the Brillouin zone boundary, 
the magnon modes involve tilting and precession 
for the atomic magnetic moments 
of only one or only the other Ru sublattice.
Conversely, in the limit $ \vec k \to \Gamma$ both sublattices are equally affected,
eventually realizing a two-fold degenerate Goldstone mode
which costs zero-energy when neglecting magneto-crystalline anisotropy (MCA),
and in which the N\'{e}el vector precesses while remaining straight
(Fig.~\ref{fig:eigenmodes limits for rutile altermagnets}). 

Away from these two extremes, i.e., for generic $\vec k$ points, 
the magnitude of tilt away from ground state magnetization direction
for the two antiparallel Ru sublattices,
i.e. their involvement in hosting the magnon excitation,
varies continuously.
In the following, we quantify it as the ratio 
between the smaller and the larger tilt,
which lies within the interval [0,1]. 
In other words, for one magnon chirality it is the ratio 
between the deviation from the ground state magnetization direction of Ru$_1$ atoms,
compared to the deviation of Ru$_2$ atoms; and vive-versa for the other magnon chirality.

We note that, although the amplitude ratio varies throughout the Brillouin zone (Fig.~\ref{fig:dispersion}c), 
the net angular momentum carried by each magnon is either $+1$ or $-1$ along the $z$-direction, 
depending only on the chirality of the magnon branch
and not on $\vec{k}$. 
Additional information on the calculation of the magnon band structure 
can be found in Appendix~\ref{sec:app_dispersion}.

\section{Simulation Results}
\label{sec:results} 
In order to account for a thermal gradient,
in the following we resort to atomistic spin dynamics simulations 
based on the time-evolution of the stochastic Landau-Lifshitz-Gilbert (sLLG) 
equation of motion \cite{Landau1935,Gilbert2004,Nowak2007}.
This describes the dynamics of the Ru magnetic moments coupled to a heat bath (see Appendix~\ref{sec:app_methods}, for details). 
In particular, a thermal gradient is simulated 
by assuming that each moment,
within a large ensemble of replicas for the RuO$_2$ unit cell,
is coupled to its own heat bath
with a fixed temperature
which varies as a function of position~\cite{Hinzke2011,Kong2013,Ritzmann2014,Kehlberger2015,Selzer2016,Mook2019,Donges2020,Weissenhofer2022}. 

As further discussed below,
magnons will in general flow from the hot to the cold terminals,
maintaining a steady current regime
across the two heat reservoirs.

In this case-study we examine this effect
as a function of both changing the steepness of the thermal gradient,
and its direction relative to the crystal lattice.
As a first temperature profile we consider the limit case 
of a very abrupt temperature step, 
with a sudden drop in temperature between hot and cold region.
We then repeat simulations for the more realistic case
of a gradual, linear temperature slope
spanning many unit cells.

We consider the direction of these temperature profiles
both along the [100] and along the [110] axis of the lattice,
which correspond, respectively, to energy-degenerate 
or to energy-split magnon dispersions in reciprocal space (see Fig.~\ref{fig:dispersion}).

In all these cases
we examine in particular the time-average 
of net magnetization per RuO$_2$ unit cell
at a given temperature,
i.e. within the same layer
perpendicular to the direction of the chosen thermal profile.
In general, this will be non-zero
when the tilt of either Ru atoms 
away from $\pm$z ground state direction
is inequivalent. 
We then also estimate the similarly layer-resolved
longitudinal spin current,
which can again be non-zero
due to the flow from hot to cold regions
of more magnons with a given chirality
than with the opposite one.

\begin{figure}
    \centering
    \includegraphics[scale=1.0]{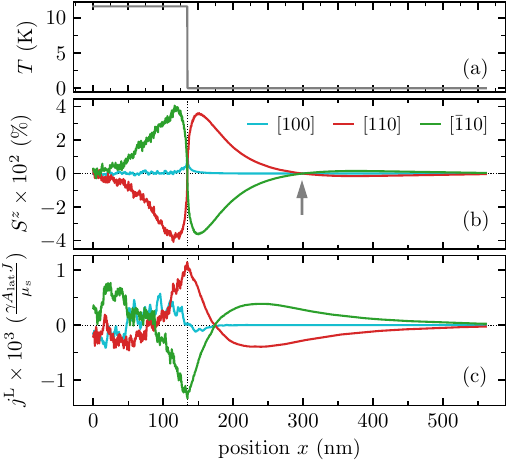}
    \caption{Simulated abrupt temperature profile (top panel a), local magnetic moment as a percentage of the ground state magnetic moment (central panel b) and longitudinal spin current (bottom panel c), for different crystallographic directions (color-coded). The grey arrow marks the sign switch of the emergent magnetization. 
    }
    \label{fig:SSE_step_0.001}
\end{figure}

We begin with results for the thermal profile
characterized by an abrupt temperature step (Fig.~\ref{fig:SSE_step_0.001}). 
When this thermal profile is aligned along the direction
corresponding to the maximal energy splitting
between the two magnon eigenmodes,
i.e. along the $[110]$ direction (or conversely, along $[\bar{1}10]$),
we observe the largest unbalance
in the flow of magnons propagating in the same hot-to-cold direction,
but with opposite chirality.
As a result, we obtain a numerical demonstration 
of a magnon-mediated spin-Seebeck effect \cite{Kehlberger2015,Uchida2010},
now associated with anisotropic scalar Heisenberg interactions
and altermagnetic low-symmetry of bulk RuO$_2$,
rather than  higher order exchange terms
such as e.g. DMI, 
which scale in magnitude with the strength of SOC.

We observe peaks in the spin accumulation,
i.e. the net magnetization per unit cell,
in the vicinity of the temperature step
(Fig.~\ref{fig:SSE_step_0.001}, panel b). 
These peaks get reproduced with opposite sign
upon rotating the thermal profile from $[110]$ to $[\bar{1}10]$ direction.

On the other hand, if the direction of the temperature profile
is aligned with any nodal plane for the magnon band structure,
such as e.g. with the $[100]$ direction,
we still have a magnon current from hot to cold regions,
but no transport of angular momentum.
This is a consequence of populating the Bose-Einstein statistics
with equal number of magnons with opposite chirality,
leading to zero net angular momentum current 
and zero emerging net magnetization per unit cell
along the whole thermal profile.

The above features, 
within numerical accuracy and statistics of the sLLG treatment,
are in accordance with
the time reversal and four-fold space rotation symmetries
of altermagnetic RuO$_2$. 

Returning to the case of the abrupt thermal profile (Fig.~\ref{fig:SSE_step_0.001})
we also observe a sign change in the emergent magnetization 
at a certain distance $\Delta = x-x_0$ from the temperature step at $x_0$.
This feature occurs at the same distance when simulations are performed along orthogonal directions e.g. 
along $[110]$ and $[\bar{1}10]$, which 
both correspond to thermal gradient orientations 
leading to the largest Seebeck effect.

We interpret it as evidence of competition between the two magnon 
bands, which coexist but contribute to a different extent and with opposite sign to the net spin current. 
In terms of Boltzmann transport theory 
we have

\begin{equation}
    S^z(x)= \sum_{\vec{k}}
    n^\beta_{\vec{k}}(x)- n^\alpha_{\vec{k}}(x)
    ,
\end{equation}

where $\alpha$ and $\beta$ denote the left- and right-handed magnon chirality, 
and $n^{\alpha(\beta)}_{\vec{k}}(x)$ is the position-dependent number of magnons. 

In this first case of an abrupt thermal gradient, we obtain $
n^{\alpha(\beta)}_{\vec{k}}(x)= n^{\alpha(\beta)}_{\vec{k}}(x_0)e^{-(x-x_0)/\lambda^{\alpha(\beta)}_{\bm{k}}}
$,
i.e. an exponential decay in the number of magnons 
as a function of distance from the abrupt temperature drop at $x_0$,
with chirality-dependent decay length $\lambda^{\alpha(\beta)}_{\vec{k}}$. 

As shown in Appendix~\ref{sec:app_analytic}, 
this formula qualitatively reproduces the features 
of the emergent magnetization
depicted in Fig~\ref{fig:SSE_step_0.001}b 
from numerical solution of the stochastic LLG equation of motion.
In particular, the key parameters 
controlling the sign and magnitude of the emergent magnetization 
are the occupation number 
(which itself depends on the temperature at each $x$), 
the decay length 
and the chirality-dependent spin angular momentum
from the magnon bandstructure. 

Within the same framework of Boltzmann transport theory,
an expression for the spin current, 
\begin{equation}
\bm{j}(x) = \sum_{\vec{k}}
n^\beta_{\vec{k}}(x)\bm{v}^\beta_{\bm{k}} - n^\alpha_{\vec{k}}(x)\bm{v}^\alpha_{\bm{k}}
,    
\end{equation}
in terms of number of magnons
and their group velocity
$
 \bm{v}^{\alpha(\beta)}_{\bm{k}}
 =
 \partial \omega^{\alpha(\beta)}_{\bm{k}}/ \partial \vec{k}$ 
for each chirality
can also be introduced. 
This quantity is not 
explicitly computed 
within atomistic spin dynamics calculations,
which only provide the values 
of an ensemble of atomic moments $\vec{S}_{is} (t)$
over the chosen  
interval of time.

Following the template of 
Refs.~\cite{Kajiwara2010,Ritzmann2015} for ferromagnets, 
we therefore introduce the following quantity,
\begin{align}
    \label{eq:spin_current_sim}
    j^\mu(\vec{r}_i)
    &
    =
    -
    \frac{\gamma}{\mu_\mathrm{s}}
    A_\mathrm{lat}
    J
    \big(
    \langle
    \vec{S}_{i\mathrm{a}}
    \times
    \vec{S}_{i+1,\mathrm{a}}
    \rangle
    +
    \langle
    \vec{S}_{i\mathrm{b}}
    \times
    \vec{S}_{i+1,\mathrm{b}}
    \rangle
    \big)
    \cdot
    \vec{e}_\mu,
\end{align}
where the average $\braket{\ldots}$ is performed over time and $J:=J_{i,i+1}^\mathrm{aa}=J_{i,i+1}^\mathrm{bb}$, with $a,b$ being sublattice indices.
Eq.~\eqref{eq:spin_current_sim} describes the transfer of the $\mu$-th component of the spin angular momentum from the unit cell $i$ (located at $\vec{r}_i$) to the adjacent unit cell $j=i+1$ and can thus be used to estimate the spin current along different spatial directions $\vec{r}_{ij}=\vec{r}_{j}-\vec{r}_{i}$. Note that we have neglected here the intersublattice correlations $\langle    \vec{S}_{is}    \times    \vec{S}_{i+1,s'\neq s}    \rangle$, which we found to be much smaller than the intrasublattice correlations. A detailed discussion as well as the derivation of Eq.~\eqref{eq:spin_current_sim} can be found in Appendix~\ref{sec:app_spin_current}.

Results from this equation for the spin current,
selecting in particular
the $z$ spin-component of the longitudinal current $j^\mathrm{L}$,
are shown in Fig.~\ref{fig:SSE_step_0.001}c for the abrupt thermal drop profile along different crystallographic directions.
In the product between statistically sampled moments at positions $\vec r_i$ and $\vec{r}_{i+1}$, the effect of stochastic fluctuations
gets magnified leading to more noisy outcome,
more pronounced in the high temperature regime.
We also expect that $j^\mathrm{L}$ 
should average to zero over the whole $x$ range
when the thermal profile is aligned along the $[100]$ direction; 
but this appears to require an increase in the system size and/or the sLLG simulation time
beyond the scope of this work.

For the $[110]$ and $[\bar{1}10]$ orientations of the thermal profile,
the spin current has an extremum directly at the step.
Its sign follows from the sign of emergent magnetization in Fig.~\ref{fig:SSE_step_0.001}b,
due to the propagation of the current always from hot to cold \cite{Cui2023}.
Same as $S^z$, the longitudinal spin current shows a flip in sign, 
but closer to the thermal drop at $x_0$. 
We obtain three extrema of $j^\mathrm{L}$, 
same as the number of zero crossing of $S^z$. 
This outcome for a steady-state solution
can be obtained 
from Boltzmann transport theory within the constant relaxation time approximation,
which prescribes $-\partial_x j^\mathrm{L} \propto S^z$
(Appendix~\ref{sec:app_analytic}).

\begin{figure}
    \centering
    \includegraphics[scale=1.0]{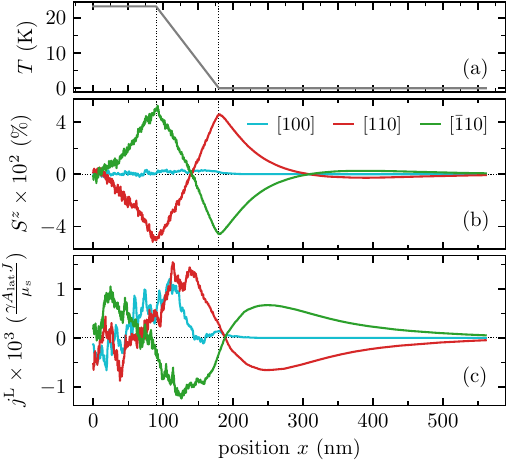}
    \caption{Simulated linear temperature gradient profile (top panel a), local magnetic moment (central panel b) and longitudinal spin current (bottom panel c), again for different crystallographic directions (color-coded).}
    \label{fig:SSE_temp_grad}
\end{figure}

We then repeat calculations
for a more realistic setup
where the two heat reservoirs are connected by a thermal gradient
with linear slope over a finite thickness,
(Fig.~\ref{fig:SSE_temp_grad}a).
The emergent $z$-magnetization reproduces some of the features
of the previous scenario of an abrupt drop in temperature
(Fig.~\ref{fig:SSE_step_0.001}),
such as the relationship between sign of $S^z$
and direction of the thermal profile.

The main difference is that $S^z$ changes linearly 
within the region with the linear temperature gradient,
with peaks exactly at the kinks of the temperature profile
(Fig.~\ref{fig:SSE_temp_grad}b).
Qualitatively the position and sign of these peaks can be understood
using linear spin wave theory 
and Boltzmann transport theory. 
Within this framework, the spin current is proportional 
to the temperature gradient 
$j^\mathrm{L}= \sigma^\mathrm{L} \partial_x T$, 
with $\sigma^\mathrm{L}$ being the longitudinal component 
of the magnon spin conductivity \cite{Cui2023}, 
and again $-\partial_x j^\mathrm{L} \propto S^z$.
Combining the two expressions,
we obtain $S^z \propto - \sigma^\mathrm{L}\partial_x^2 T$. 
This means that the emergence of a finite $S^z$ 
requires non-linearities of the temperature profile. 

However, while this simple formula is helpful in predicting 
whether a finite magnetization emerges at all, 
it cannot explain its length scales.
Quantitative values require numerical sLLG simulations.

We finally investigate 
also the generation of transverse spin currents $j^\mathrm{T}$ via the spin Nernst effect. 
In a recent work \cite{Cui2023} it has been predicted that strength of the spin Nernst effect in altermagnets can be significantly larger than in other collinear antiferromagnets, 
where it emerges due to nonzero Berry curvature \cite{Cheng2016,Park2020}. 

\begin{figure}[ht!]
    \centering
    \includegraphics[scale=1.0]{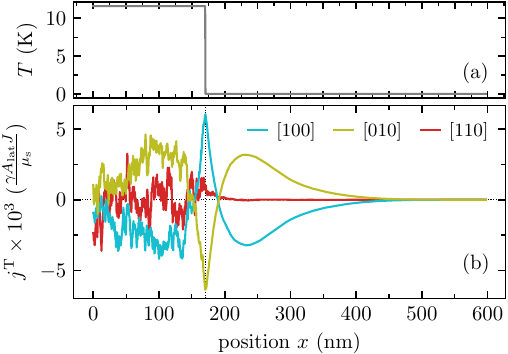}
    \caption{Simulated abrupt temperature profile (top panel a), and transversal spin current (bottom panel b), for different crystallographic directions (color-coded).}
    \label{fig:SNE_step_0.001}
\end{figure}

In Fig.~\ref{fig:SNE_step_0.001} we depict results for the transverse spin current obtained using Eq.~\eqref{eq:spin_current_sim} for the same temperature step profile discussed above. In line with the predictions of \cite{Cui2023}, our simulations clearly show that only magnon currents traveling along the $[100]$ and $[010]$ directions lead to the emergence of a transverse spin current. The fluctuating values of $j^\mathrm{T}$ for the $[110]$ case can be attributed to the aforementioned slow convergence of this quantity. This is underlined by calculations within Boltzmann transport theory (shown in Appendix~\ref{sec:app_analytic}), which support our numerical findings for the $[100]$ and $[010]$ cases and predict a vanishing transverse spin current for the $[110]$ case. Remarkably, the spatial profile of the transverse spin current is found to be very similar to the one of the longitudinal spin current, with a magnitude that is roughly a factor two smaller.

Note that in our simulations the emerging transverse spin current cannot lead to a finite spin accumulation, due to the lack of translational symmetry breaking along the transverse direction. This is because we use periodic boundary conditions along $y,z$, i.e. perpendicularly to the direction $x$ of the temperature profile, in order to reduce possible numerical artifact due to finite simulation size. The simulation of real 2D (or even 3D) systems with diameters of several hundred nanometres (containing more than $10^8$ Ru moments) is beyond the scope of this work. Real systems, of course, are finite and thus we expect that the transverse spin current results in a spin accumulation -- a finite net magnetization -- with the same spatial profile (with opposite sign) at the opposing edges. We propose that this spin accumulation could be experimentally detected using, e.g., magneto-optical measurement analogous to the detection of spin \cite{Stamm2017} or orbital Hall effects \cite{Choi2023,Lyalin2023}.

\section{\label{sec: conclusions}Conclusion}

We predict that spin-split antiferromagnets (or altermagnets) 
host magnon currents traveling from the hot to the cold reservoirs.
Specific to this class of materials,
the non-degenerate, chirality-dependent energy splitting of the magnon band structure,
which is present even in the absence of higher order Heisenberg Hamiltonian terms
(such as e.g. Dzialoshinskii-Moriya interactions),
should lead to different propagation of magnons with anti-parallel spin angular momentum
depending on the direction of the thermal gradient
with respect to the crystal,
and in particular to the anisotropy in the non-relativistic (scalar) Heisenberg exchange interactions.

This outcome is consistent with
the symmetry of the spin currents and the ensuing spin accumulation 
according to spin-group analysis,
and it has been both quantitatively explored
through numerical stochastic Landau-Lifshitz-Gilbert (sLLG) calculations
based on parameters from ab initio electronic structure theory,
and through Boltzmann transport theory in the constant relaxation time approximation.

In particular, we provide an estimate of magnon-mediated Seebeck and Nernst effects 
associated with the non-relativistic component of the non-degenerate spin-wave dispersion in altermagnets,
in the particular case of RuO$_2$, chosen here as a reference example for altermagnets. Our conclusions also apply, qualitatively, to other materials with alike unit cell geometry and magnetic order, such as e.g. MnF$_2$, FeF$_2$, CoF$_2$, and to cases with more than two sublattices with antiparallel magnetic moments and/or with a different than four-fold rotation symmetry, e.g., Fe$_3$O$_4$ or MnTe. 

Since the sign and spatial profile of the spin accumulation generated by the emergent magnon currents are fundamentally linked to the altermagnetic spin-group symmetry, we propose that experimental detection of the ensuing spin accumulation, e.g. by magneto-optical measurements, could be a feasible method for an indirect identification of materials, particularly insulators, as altermagnetic.

\acknowledgments
The authors thank S. Mankovsky for valuable discussions.
M.W. acknowledges financial support from the German Research Foundation (Deutsche Forschungsgemeinschaft) through CRC/TRR 227 "Ultrafast Spin Dynamics".
A.M. acknowledges partial financial support 
from the Czech Science Foundation (GA CR) grant 23-04746S,
and from a "long-term conceptual development of a research organization" (DKRVO)
grant from the University of West Bohemia in Plzen.
The computations were enabled by resources provided by the National Academic Infrastructure for Supercomputing in Sweden (NAISS) at NSC Linköping partially funded by the Swedish Research Council through grant agreement no. 2022-06725.

\appendix

\section{Calculation of magnon dispersion in altermagnets}
\label{sec:app_dispersion}
In this section we derive the magnon band structure for a general altermagnet using second quantization. 
We start from a two-sublattice Heisenberg Hamiltonian,
\begin{align}
    \label{eq:app_Hamiltonian}
    \hat{\mathcal{H}} = - \frac{1}{S^2} \sum_{ij} \sum_{ss'} J^{ss'}_{ij} \hat{\bm{s}}_{is} \cdot \hat{\bm{s}}_{js'},
\end{align}
with $\hat{\bm{s}}_{is}$ denoting spin operators with spin quantum number $S$ and $J^{ss'}_{ij}$ the (scalar) exchange coupling between the magnetic moments for the atoms 
respectively at $\vec r_i + \vec b_{s}$
and at $\vec r_j + \vec b_{s'}$, 
where $\vec r_i$ ($\vec r_j$) is the location of the unit cell within a periodic lattice,
and $\vec b_{s}$ ($\vec b_{s'}$) the sublattice position within it.

We follow the usual scheme for the derivation of an antiferromagnetic magnon dispersion relation based on the Holstein-Primakoff transformation \cite{Holstein1940}, which up to first order in magnon variables is given by
\begin{align}
    \begin{split}
    \hat{s}^+_{i\mathrm{a}}&\approx\sqrt{2S}\hat{a}_i,
    \hspace{1em}
    \hat{s}^-_{i\mathrm{a}}\approx\sqrt{2S}\hat{a}^\dagger_i,
    \hspace{1em}
    \hat{s}^z_{i\mathrm{a}}=S-\hat{a}^\dagger_i\hat{a}_i
    \\
    \hat{s}^+_{i\mathrm{b}}&\approx\sqrt{2S}\hat{b}^\dagger_i,
    \hspace{1em}
    \hat{s}^-_{i\mathrm{b}}\approx\sqrt{2S}\hat{b}_i,
    \hspace{1em}
    \hat{s}^z_{i\mathrm{b}}=-S+\hat{b}^\dagger_i\hat{b}_i,    
    \end{split}
\end{align}
with $\hat{s}^\pm_{is}=\hat{s}^x_{is}\pm i\hat{s}^y_{is}$, and the magnon creation ($\hat{a}^\dagger_i,\hat{b}^\dagger_i$) and annihilation operators ($\hat{a}_i,\hat{b}_i$).

Inserting into the Hamiltonian~\eqref{eq:app_Hamiltonian} and Fourier transformation (with $
\hat{a}^{(\dagger)}_i
=
\sqrt{1/N}
\sum_{\vec{k}} \exp(-i \vec{k} \cdot \vec{r}_i) \hat{a}^{(\dagger)}_{\vec{k}}
$ 
and analogous for $\hat{b}_i^{(\dagger)}$) yields $\hat{\mathcal{H}}\approx\mathrm{const.} + \hat{\mathcal{H}}^\mathrm{mag}$, with
\begin{align}
    &\hat{\mathcal{H}}^\mathrm{mag}
    =
    \sum_{\vec{k}}
    \begin{pmatrix}
        \hat{a}^\dagger_{\bm{k}} & \hat{b}_{\bm{k}}
    \end{pmatrix}
     \mathcal{H}_{\bm{k}}
     \begin{pmatrix}
        \hat{a}_{\bm{k}} \\ \hat{b}^\dagger_{\bm{k}}
    \end{pmatrix}
    ,
    \\
    &\mathcal{H}_{\bm{k}}
    =
    -
    \frac{2}{S}
    \begin{pmatrix}
        \tilde{J}^{\mathrm{aa}}_{\bm{k}}
        -
        \tilde{J}^{\mathrm{aa}}_{\bm{0}}
        +
        \tilde{J}^{\mathrm{ab}}_{\bm{0}}
        &
        \tilde{J}^{\mathrm{ab}}_{\bm{k}}
        \\
        \tilde{J}^{\mathrm{ab}}_{\bm{k}}
        &
        \tilde{J}^{\mathrm{bb}}_{\bm{k}}
        -
        \tilde{J}^{\mathrm{bb}}_{\bm{0}}
        +
        \tilde{J}^{\mathrm{ab}}_{\bm{0}}
    \end{pmatrix}
    .
\end{align}
Note that we have introduced the Fourier transformed exchange coupling constants $\tilde{J}^{ss'}_{\bm{k}} = \sum_j \exp(i \bm{k} \cdot \bm{r}_j) J_{ij}^{ss'}$.

Next, we perform a bosonic Bogoliubov–Valatin transformation \cite{Bogoljubov1958,Valatin1958} to diagonalize the Hamiltonian. We introduce new magnon variables ($\hat{\alpha}^{(\dagger)}_{\bm{k}},\hat{\beta}^{(\dagger)}_{\bm{k}}$) via
\begin{align}
    \hat{a}_{\bm{k}}
    &=
    u_{\bm{k}}
    \hat{\alpha}_{\bm{k}}
    -
    v_{\bm{k}}
    \hat{\beta}^\dagger_{\bm{k}}
    \\
    \hat{b}^\dagger_{\bm{k}}
    &=
    u_{\bm{k}}
    \hat{\beta}^\dagger_{\bm{k}}
    -
    v_{\bm{k}}
    \hat{\alpha}_{\bm{k}},
\end{align}
where the real coefficients $u_{\bm{k}}$ and $v_{\bm{k}}$ fulfill $u_{\bm{k}}^2-v_{\bm{k}}^2=1$ in order to conserve the canonical commutation relations.
From this we get the diagonalized magnon Hamiltonian
\begin{equation}
    \hat{\mathcal{H}}^\mathrm{mag}
    = 
    \sum_{\vec{k}}
    \hbar\omega_{\bm{k}}^\alpha\hat{\alpha}^\dagger_{\bm{k}}\hat{\alpha}_{\bm{k}}
    +
    \hbar\omega_{\bm{k}}^\beta\hat{\beta}^\dagger_{\bm{k}}\hat{\beta}_{\bm{k}}
\end{equation}
with the magnon frequencies
\begin{align}
    \omega^\alpha_{\bm{k}}
    &=
    \frac{
    \omega^\mathrm{a}_{\bm{k}}-\omega^\mathrm{b}_{\bm{k}}
    }{2}
    +
    \sqrt{
    \bigg(
   \frac{
   \omega^\mathrm{a}_{\bm{k}}+\omega^\mathrm{b}_{\bm{k}}
    }{2}
    -
    \frac{
    S\tilde{J}^\mathrm{ab}_{\vec{0}}
    }{\hbar}
    \bigg)^2
    -
    \bigg(
    \frac{
    S\tilde{J}^\mathrm{ab}_{\vec{k}}
    }{\hbar}
    \bigg)^2
    }
    \\
    \omega^\beta_{\bm{k}}
    &=
    \frac{
    \omega^\mathrm{a}_{\bm{k}}-\omega^\mathrm{b}_{\bm{k}}
    }{2}
    -
    \sqrt{
    \bigg(
   \frac{
   \omega^\mathrm{a}_{\bm{k}}+\omega^\mathrm{b}_{\bm{k}}
    }{2}
    -
    \frac{
    S\tilde{J}^\mathrm{ab}_{\vec{0}}
    }{\hbar}
    \bigg)^2
    -
    \bigg(
    \frac{
    S\tilde{J}^\mathrm{ab}_{\vec{k}}
    }{\hbar}
    \bigg)^2
    }.
\end{align}
Here, we introduced $\omega^a_{\bm{k}}=2(\tilde{J}^\mathrm{aa}_{\bm{0}} -    \tilde{J}^\mathrm{aa}_{\bm{k}})/(S\hbar)$ and $\omega^b_{\bm{k}}=2(\tilde{J}^\mathrm{bb}_{\bm{0}} -    \tilde{J}^\mathrm{bb}_{\bm{k}})/(S\hbar)$. 
For zero intersublattice coupling one recovers two decoupled ferromagnetic dispersion relations 
$\omega^\alpha_{\bm{k}}=\omega^\mathrm{a}_{\bm{k}}$ and 
$\omega^\beta_{\bm{k}}=\omega^\mathrm{b}_{\bm{k}}$ 
which are separately hosted on the distinct sublattices.
Similarly, the magnon frequencies reduce to the typical antiferromagnetic magnon dispersion relation $\omega^\alpha_{\bm{k}}  =2\sqrt{(\tilde{J}^\mathrm{ab}_{\bm{0}})^2-(\tilde{J}^\mathrm{ab}_{\bm{k}})^2}/(S\hbar)$ and $\omega^\beta_{\bm{k}}  = -    2\sqrt{(\tilde{J}^\mathrm{ab}_{\bm{0}})^2-(\tilde{J}^\mathrm{ab}_{\bm{k}})^2}/(S\hbar)$ in the absence of intrasublattice coupling.
The magnon eigenmodes for RuO$_2$ are depicted in Fig.~\ref{fig:dispersion}b.

Unlike the magnons in the original ($\hat{a}^{(\dagger)}_{\bm{k}},\hat{b}^{(\dagger)}_{\bm{k}}$) basis, which only have a finite amplitude at one of the sublattices, the magnons in the new ($\hat{\alpha}^{(\dagger)}_{\bm{k}},\hat{\beta}^{(\dagger)}_{\bm{k}}$) basis are characterized by a tilting of the moments in both sublattices,
\begin{align}
    \begin{split}
    &
    \Delta \hat{s}^z_\mathrm{a}
    = 
    \sum_{i} S-\hat{S}^z_{i\mathrm{a}}
    \\
    &=    
    \sum_{\bm{k}}
    u_{\bm{k}}^2
    \hat{\alpha}^\dagger_{\bm{k}}
    \hat{\alpha}_{\bm{k}}
     +
    v_{\bm{k}}^2
    (
    \hat{\beta}^\dagger_{\bm{k}}
    \hat{\beta}_{\bm{k}}
    +
    1
    )
    -
    u_{\bm{k}}
    v_{\bm{k}}
    (
    \hat{\alpha}^\dagger_{\bm{k}}
    \hat{\beta}^\dagger_{\bm{k}}
    +
    \hat{\alpha}_{\bm{k}}
    \hat{\beta}_{\bm{k}}
    ),
    \\    
    \end{split}
    \\
    \begin{split}
    &\Delta \hat{s}^z_\mathrm{b}
    = 
    \sum_{i} S+\hat{S}^z_{i\mathrm{b}}
    =    \\
    &
    \sum_{\bm{k}}
    u_{\bm{k}}^2
    \hat{\beta}^\dagger_{\bm{k}}
    \hat{\beta}_{\bm{k}}
     +
    v_{\bm{k}}^2
    (
    \hat{\alpha}^\dagger_{\bm{k}}
    \hat{\alpha}_{\bm{k}}
    +
    1
    )
    -
    u_{\bm{k}}
    v_{\bm{k}}
    (
    \hat{\alpha}^\dagger_{\bm{k}}
    \hat{\beta}^\dagger_{\bm{k}}
    +
    \hat{\alpha}_{\bm{k}}
    \hat{\beta}_{\bm{k}}
    ).     
    \end{split}
\end{align}
From that we get that the sublattice amplitudes for the $\alpha$ magnon branch are $\braket{0|\hat{\alpha}_{\bm{k}}\Delta \hat{s}^z_{a}\hat{\alpha}^\dagger_{\bm{k}}|0}=u_{\bm{k}}^2+v_{\bm{k}}^2$ and $\braket{0|\hat{\alpha}_{\bm{k}}\Delta \hat{s}^z_{b}\hat{\alpha}^\dagger_{\bm{k}}|0}=2v_{\bm{k}}^2$, while for the $\beta$ magnon branch they are $\braket{0|\hat{\beta}_{\bm{k}}\Delta \hat{s}^z_{a}\hat{\beta}^\dagger_{\bm{k}}|0}=2v_{\bm{k}}^2$ and $\braket{0|\hat{\beta}_{\bm{k}}\Delta \hat{s}^z_{b}\hat{\beta}^\dagger_{\bm{k}}|0}=u_{\bm{k}}^2+v_{\bm{k}}^2$. Here, $\ket{0}$ denotes the ground state of the system. Henceforth, the ratio $R_{\bm{k}}$ of the respective smaller and larger magnon amplitude at each sublattice is given by $R_{\bm{k}} = 2v^2_{\bm{k}}/(u^2_{\bm{k}}+v^2_{\bm{k}})=2v^2_{\bm{k}}/(1+2v^2_{\bm{k}}) \leqslant 1$. This quantity calculated for the case of RuO$_2$ is shown in Fig.~\ref{fig:dispersion}c.

The operator for the total $z$-magnetization in the new basis set is given by the difference in magnon numbers in each branch before and after the Bogoliubov–Valatin transformation,
\begin{align}
    \begin{split}
    \hat{s}^z &
    = 
    \sum_{i} \hat{S}^z_{ia} + \hat{S}^z_{ib}
    =
    \sum_{\bm{k}}
    \hat{b}^\dagger_{\bm{k}}\hat{b}_{\bm{k}}
    -
    \hat{a}^\dagger_{\bm{k}}\hat{a}_{\bm{k}}
    =
    \sum_{\bm{k}} 
    \hat{\beta}^\dagger_{\bm{k}}\hat{\beta}_{\bm{k}}
    -
    \hat{\alpha}^\dagger_{\bm{k}}\hat{\alpha}_{\bm{k}}.
    \end{split}
\end{align}
Consequently, the $\alpha$ and $\beta$ magnons carry opposite spin angular momentum along the $z$-direction, since $\braket{0|\hat{\alpha}_{\bm{k}}\hat{s}^z\hat{\alpha}^\dagger_{\bm{k}}|0}=-1$ and $\braket{0|\hat{\beta}_{\bm{k}}\hat{s}^z\hat{\beta}^\dagger_{\bm{k}}|0}=1$.

\section{Boltzmann transport theory for magnon currents and spin accumulation}
\label{sec:app_analytic}

Within our approach, the $z$ component of the magnetization $S^z=\langle \hat{s}^z \rangle$ is given by the difference of the total magnon numbers in each branch,
\begin{equation}
     S^z
     =
    \sum_{\bm{k}} 
    \langle\hat{\beta}^\dagger_{\bm{k}}\hat{\beta}_{\bm{k}}\rangle 
    -
    \langle\hat{\alpha}^\dagger_{\bm{k}}\hat{\alpha}_{\bm{k}}\rangle 
    =
    \sum_{\bm{k}} n^\beta_{\bm{k}} - n^\alpha_{\bm{k}}
    \label{eq:app_S_z_average}
    .
\end{equation}

To describe spatially non-homogeneous and possibly time-dependent magnon occupations, we use Boltzmann equations within the relaxation time ansatz,
\begin{align}
\frac{\partial n^\sigma_{\bm{k}}(\bm{r},t)}{\partial t}
+
\frac{\partial n^\sigma_{\bm{k}}(\bm{r},t)}{\partial \bm{r}}
\cdot
\frac{\partial \omega^\sigma_{\bm{k}}}{\partial \bm{k}}
&=
-
\frac{
n^\sigma_{\bm{k}}(\bm{r},t)
-
f^\sigma_{\bm{k}}(\bm{r},t)
}{\tau^\sigma_{\bm{k}}},
\label{eq:Boltzmann_eq}
\end{align}
with $\sigma\in [\alpha,\beta]$. Here, $\partial \omega^\sigma_{\bm{k}}/\partial \bm{k}=\bm{v}_{\bm{k}}$ and $\tau^\sigma_{\bm{k}}$ are the branch- and $\bm{k}$-dependent magnon velocities and lifetimes, respectively, and $f^\sigma_{\bm{k}}(\bm{r},t)$ is the equilibrium magnon occupation.

In steady state and for the current along the $x$-direction, we get that 
\begin{equation}
\frac{\partial n^\sigma_{\bm{k}}(x)}{\partial x}
=
-
\frac{n^\sigma_{\bm{k}}(x)
-
f^\sigma_{\bm{k}}(x)}{\bm{v}^\sigma_{\bm{k}}\tau^\sigma_{\bm{k}}}
.
\end{equation}

Hereinafter, we solve this equation for the case of a temperature step at $x=0$ that connects two regions of constant temperature. In this case, we have a constant equilibrium magnon occupations $f^\sigma_{\vec{k}}(x<0) = c^\sigma_{\vec{k}}$ and $f^\sigma_{\vec{k}}(x\geqslant 0) = d^\sigma_{\vec{k}}$ in the two regions. A simple calculation yields that the occupation number of any magnon with $v>0$ has the solution
\begin{equation}
n^{\sigma,v>0}_{\vec{k}}(x)
=
\begin{cases}
c^\sigma_{\vec{k}} & \mathrm{for}\ x < 0
\\
(c^\sigma_{\vec{k}}-d^\sigma_{\vec{k}})
e^{-x/(\bm{v}^\sigma_{\bm{k}}\tau^\sigma_{\bm{k}})}
+
d^\sigma_{\vec{k}}
& \mathrm{for}\ x \geq 0
\end{cases}
\end{equation}
and, analogously, the occupation number of magnons with $v<0$ follows
\begin{equation}
n^{\sigma,v<0}_{\vec{k}}(x)
=
\begin{cases} 
(d^\sigma_{\vec{k}}-c^\sigma_{\vec{k}})
e^{-x/(\bm{v}^\sigma_{\bm{k}}\tau^\sigma_{\bm{k}})}
+
c^\sigma_{\vec{k}}
& \mathrm{for}\ x < 0
\\
d^\sigma_{\vec{k}}
& \mathrm{for}\ x \geq 0.
\end{cases}
\end{equation}

\begin{figure}
    \centering
    \includegraphics{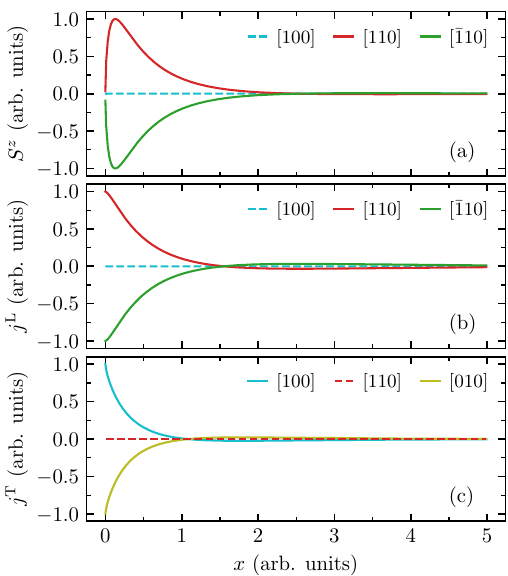}
    \caption{Spin-Seebeck and spin-Nernst effect in altermagnetic RuO$_2$ induced by a temperature step calculated by Boltzmann transport theory. Spin accumulation $S^z$ (top panel a), longitudinal spin current (central panel b) and transverse spin current (bottom panel c) are shown versus position coordinate for different crystallographic directions (color-coded).}
    \label{fig:SSE_step_analytisch}
\end{figure}

To simplify the following calculations we assume that $T(x>0)$ is zero, which implies that $d^\sigma_{\bm{k}}=0$, and compute the solution only in the region $x>0$. Inserting into Eq.~\eqref{eq:app_S_z_average} yields
\begin{align}
    \begin{split}
        S^z(x>0)
        &=
        \sum_{\bm{k}} n^{\beta,v>0}_{\vec{k}}(x)-n^{\alpha,v>0}_{\vec{k}}(x)
        \\
        &=
        \sum_{\bm{k}} 
        c^\beta_{\vec{k}}
        e^{-x/\lambda^\beta_{\bm{k}}}
        - 
        c^\alpha_{\vec{k}}
        e^{-x/\lambda^\alpha_{\bm{k}}},
    \end{split}
\end{align}
where the decay length $\lambda^\sigma_{\vec{k}}=\bm{v}^\beta_{\bm{k}}\tau^\beta_{\bm{k}}$ was introduced. To evaluate the above expression, it is necessary to know the lifetimes $\tau^\sigma_{\vec{k}}$ of the magnon modes. Following Ref.~\cite{Cui2023}, we apply a constant lifetime approximation $\tau^\sigma_{\vec{k}}\equiv\tau_0$. 
Since we want to compare with spin dynamics simulations based on the sLLG, we assume a classical equilibrium occupation $c^\sigma_{\vec{k}}=k_\mathrm{B}T/(\hbar \omega^\sigma_{\bm{k}})$ \cite{Barker2016}, rather than a Bose-Einstein distribution. The result for the emergent magnetization for different orientations of the temperature step is shown in Fig.~\ref{fig:SSE_step_analytisch}a and agrees qualitatively very well with what we obtained via simulations (see Sec.~\ref{sec:results}).

Summation over $\bm{k}$ transforms Eq.~\eqref{eq:Boltzmann_eq} to the respective continuity equations for each magnon branch,
\begin{align}
\frac{\partial }{\partial t}
N_\sigma(\bm{r},t)
+
\bm{\nabla}
\cdot
\bm{j}_\sigma(\bm{r},t)
&=
-\Gamma_\sigma(\bm{r},t),
\label{eq:con_eq_sl}
\end{align}
where $N_\sigma(\bm{r},t)=\sum_{\bm{k}}n^\sigma_{\vec{k}}(\bm{r},t)$ is the total magnon number in each branch, $\bm{j}_\sigma(\bm{r},t)=\sum_{\bm{k}}n^\sigma_{\vec{k}}(\bm{r},t)\bm{v}^\sigma_{\bm{k}}$ is the magnon current carrying $z$-component spin angular momentum and $\Gamma_\sigma(\bm{r},t)=(N_\sigma(\bm{r},t)-N^0(\bm{r},t))/\tau_0$ is a sink term. Note that the total number of magnons in the equilibrium $N^0(\bm{r},t)$ is the same for both magnon branches due to the summation over the entire Brillouin zone.

Recalling that the $z$-magnetization is determined by the difference between the respective magnon numbers in each branch -- cf.\ Eq.~\eqref{eq:app_S_z_average} -- we subtract the continuity equation for the $\alpha$ branch from the one for the $\beta$ branch, yielding
\begin{align}
\begin{split}
\frac{\partial }{\partial t}
\Big(
N_\beta(\bm{r},t)
-
N_\alpha(\bm{r},t)
\Big)
+
\bm{\nabla}
\cdot
\Big(
&\bm{j}_\beta(\bm{r},t)
-
\bm{j}_\alpha(\bm{r},t)
\Big)\\
&=
-
\frac{
N_\beta(\bm{r},t)
-
N_\alpha(\bm{r},t)
}{\tau_0}.
\end{split}
\end{align}
By that, we arrive at the continuity equation of the emergent magnetization
\begin{equation}
\frac{\partial  S^z(\bm{r},t)}{\partial t}
+
\bm{\nabla}
\cdot
\bm{j}(\bm{r},t)
=
-
\frac{
 S^z(\bm{r},t)
}{\tau_0},
\label{eq:con_eq_altmag}
\end{equation}
where we introduced the spin current as
\begin{equation}
    \bm{j}(\bm{r},t)
    =
    \sum_{\bm{k}}
    n^\beta_{\vec{k}}(\bm{r},t)\bm{v}^\beta_{\bm{k}}
    -
    n^\alpha_{\vec{k}}(\bm{r},t)\bm{v}^\alpha_{\bm{k}}.
\end{equation}
Unsurprisingly, it is given by the sum of the magnon currents $\bm{j}_\mathrm{s}^\sigma(\bm{r},t)= \sum_{\bm{k}}   n^\sigma_{\vec{k}}(\bm{r},t)\bm{v}^\sigma_{\bm{k}}$ of each sublattice multiplied by the respective spin angular momentum along the $z$-direction of $\pm1$. In Fig.~\ref{fig:SSE_step_analytisch}b and c we demonstrate that the longitudinal and transverse magnonic spin currents are again in good qualitative agreement with the simulation results shown in Sec.~\ref{sec:results}.

For steady-state ($\partial S^z(\vec{r},t)/\partial t =0$), we get 
\begin{equation}
   \bm{\nabla}
\cdot
\bm{j}(\bm{r})
=
-
\frac{
 S_z(\bm{r})
}{\tau_0}.
\end{equation}
While we found this relation to be not exactly fulfilled in the simulations (see Sec.~\ref{sec:results}) -- most likely due to shortcomings of the constant lifetime approximation -- it nonetheless provides a useful relation between between the magnonic spin current and the emergent magnetization.

\section{Simulation methods}
\label{sec:app_methods}

We perform atomistic spin dynamics simulations based on the
stochastic Landau-Lifshitz-Gilbert (sLLG) equation of motion
for a sufficiently big ensemble of atomic magnetic moments \cite{Landau1935,Gilbert2004,Nowak2007},
\begin{align}
    \frac{\partial \vec{S}_{is}}{ \partial t}
    &=
    -
    \frac{\gamma}{(1+\alpha^2)\mu_\mathrm{s}}
    \vec{S}_{is}
    \times
    (
    \vec{H}_{is}
    +
    \alpha
    \vec{S}_{is}
    \times
    \vec{H}_{is}
    ).
\end{align}

Here $\vec{S}_{is}$ denotes a unit vector in the direction of the Ru magnetic moment of sublattice $s$ in the unit cell at $i$, $\alpha$ is the Gilbert damping parameter, $\mu_\mathrm{s}$ is the saturation magnetic moment and $\gamma$ is the gyromagnetic ratio. 

This numerical approach goes beyond non-interacting spin-wave theory,
as used in previous works such as e.g. Ref.~\cite{Smejkal2023_chiral_magnons,Cui2023},
and may provide a more realistic description of experiments \cite{Elliott1969}
in terms of adiabatic magnon dispersion.
On the other hand, it neglects intrinsic damping effects due to the Stoner continuum,
which can be large in metallic magnetic materials
and would show up as a possibly shifted position of the peaks 
associated with magnon dispersion, 
and their their progressively broader FWHM,
particularly toward the boundary of the Brillouin zone.
Within the framework of the sLLG, damping is included via the Gilbert term (proportional to the Gilbert damping parameter $\alpha$), due to which each individual atomic magnetic moment 
tends to gradually spiral back toward its ground state direction.

\begin{figure}[h!]
    \centering
    \includegraphics[scale=1.0]{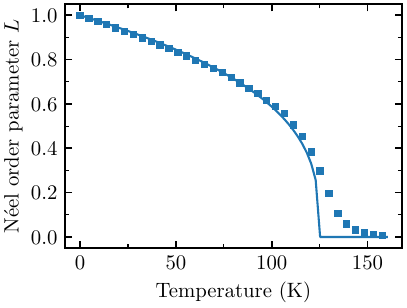}
    \caption{N{\'e}el order parameter $L$ of altermagnetic RuO$_2$ for a system consisting of $128000$ Ru moments calculated using the sLLG and spin model parameters from Ref.~\cite{Smejkal2023_chiral_magnons}. The simulation results are fitted to the expression $L(T)=(1-T/T_\mathrm{c})^{0.33}$ (shown as solid line) which yields a fitted $T_\mathrm{c}\approx\SI{125}{\kelvin}$.}
    \label{fig:Tc}
\end{figure}

Since we are only interested in steady state solutions of the sLLG, we can set $\mu_\mathrm{s}=\gamma=1$.
To the best of our knowledge, the Gilbert damping parameter for RuO$_{2}$ is unknown. That is why we choose an intermediate value of $\alpha=0.001$ for the simulations presented in Sec.~\ref{sec:results}. We have also performed simulations with $\alpha=0.01$ (not shown) and found that this only affects the magnitude and length scale of the resulting effects, while they are qualitatively the same. 
The effective field $\vec{H}_{is}=-\partial \mathcal{H}/\partial\vec{S}_{is}+\vec{\zeta}_{is}$ contains both a deterministic field that stems from the spin Hamiltonian~\eqref{eq:Hamiltonian} and a stochastic field $\vec{\zeta}_{is}$ in the form of Gaussian white noise \cite{Brown1960},
\begin{align}
    \braket{\vec{\zeta}_{is}}&=0
    ,
    \\
    \big\langle
    \vec{\zeta}_{is}(t)\big(\vec{\zeta}_{is}(0)\big)^\mathrm{T}
    \big\rangle
    &=
    2
    \frac{\alpha k_\mathrm{B}T_i \mu_\mathrm{s}}{\gamma}
    \delta_{ij}
    \delta_{ss'}
    \delta(t)
    \mathbb{1},
\end{align}
which models a coupling to a heat bath -- incorporating the interaction of localized spins with electronic and phononic degrees of freedom -- with a local temperature $T_i$.
The numerical integration of the sLLG is carried out using the Heun algorithm \cite{Nowak2007}. To reduce the computational cost we introduce an energy cutoff for the Heisenberg exchange constants such that only those with $|J^{ss'}_{ij}|\geqslant \SI{0.0625}{\milli\electronvolt}$ are included. By calculating the N{\'e}el order parameter $L=|\frac{1}{2N}\sum_i^{N} \vec{S}_{ia}-\vec{S}_{ib}|$, with $N$ being the number of unit cells and a and b denoting the Ru respective sublattices, versus temperature we estimate a critical temperature of $T_\mathrm{c}\approx\SI{125}{\kelvin}$ (see Fig.~\ref{fig:Tc}).
In our simulations of the magnon currents and the emergent magnetization we consider a system extended in the direction of the temperature step/gradient consisting of $36\times36\times1600$ unit cells (equalling to $4147200$ Ru moments). Moreover, we apply periodic boundary conditions in the directions perpendicular to the temperature step/gradient to eliminate finite-size effects.

\section{Numerical calculation of spin currents in atomistic spin dynamics simulations}
\label{sec:app_spin_current}
When doing magnon transport calculations, theoretical models based on linear response theory and Boltzmann transport theory typically only compute the spin current, rather than the associated magnetization dynamics. Atomistic spin dynamics simulations, on the other hand, are used to directly calculate the temporal evolution of the magnetic moments, which only indirectly contains information about the spin currents via a possibly arising spin accumulation. To bridge this gap, earlier works introduced formulas to obtain the spin currents in ferromagnets within continuum theory \cite{Kajiwara2010} and atomistic descriptions including only isotropic nearest neighbor Heisenberg exchange \cite{Ritzmann2015}. Hereinafter, we derive an expression to calculate the spin current for multisublattice systems such as, e.g., antiferro- and altermagnets, with arbitrary Heisenberg exchange interaction extending beyond nearest neighbors.

Neglecting losses due to Gilbert damping, the time averaged changed of spin in a single unit cell with index $i$ follows from the Landau-Lifshitz-Gilbert equation as
\begin{align}
    \label{eq:app_change_of_spin}
    \Big\langle
    \sum_s\dot{\vec{S}}_{is}
    \Big\rangle
    =
    -
    \frac{\gamma}{\mu_\mathrm{s}}
    \sum_{ss',\delta }
    J_{i,i+\delta}^{ss'}
    \Big\langle
    \vec{S}_{is}
    \times
    \vec{S}_{i+\delta, s'}
    \Big\rangle
    ,
\end{align}
with $s,s'$ being the sublattice indices. For simplicity, we consider only a 1D chain of unit cells. Generalization to 3D systems, however, is straightforward. The right-hand side of Eq.~\eqref{eq:app_change_of_spin} can be viewed as the difference between incoming ($\delta<0$) and outgoing ($\delta>0$) spin momentum. Note that the terms for $\delta=0$, i.e., within the chosen unit cell, mutually cancel due to the antisymmetry of the cross product. Thus, we can write $ \braket{\sum_s\dot{\vec{S}}_{is}} = -  (\vec{j}^\mathrm{in}_i   -   \vec{j}^\mathrm{out}_i   )/d$ with the incoming and outgoing spin currents 
\begin{align}
    \vec{j}^\mathrm{in}_i
    &=
    \frac{\gamma}{\mu_\mathrm{s}}
    d
    \sum_{ss',\delta < 0 }
    J_{i,i+\delta}^{ss'}
    \Big\langle
    \vec{S}_{is}
    \times
    \vec{S}_{i+\delta, s'}
    \Big\rangle
    ,
    \\
    \vec{j}^\mathrm{out}_i
    &=
    -
    \frac{\gamma}{\mu_\mathrm{s}}
    d
    \sum_{ss',\delta > 0 }
    J_{i,i+\delta}^{ss'}
    \Big\langle
    \vec{S}_{is}
    \times
    \vec{S}_{i+\delta, s'}
    \Big\rangle
\end{align}

 In steady-state (for which $\braket{\sum_s\dot{\vec{S}}_{is}}=0$) incoming and outgoing currents are equal, $\vec{j}^\mathrm{in}_i=\vec{j}^\mathrm{out}_i:=\vec{j}_i$, and henceforth either of the expressions above can be used to calculate the spin current $\vec{j}_i$ for arbitrary $J_{ij}^{ss'}$ and ground state configuration.

\begin{figure}
    \centering
    \includegraphics{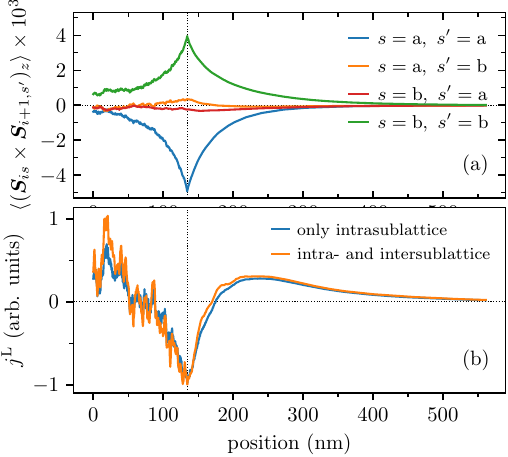}
    \caption{Panel (a): Intra- ($s=s'$) and intersublattice ($s\neq s'$) correlations for a temperature step with $T=\SI{10}{\kelvin}$ to the left of the vertical dotted line and zero to the right. Panel (b): spatial profile of the $z$-component of the longitudinal spin current calculated considering only intrasublattice correlations and both intra- and intersublattice correlations with the next unit cell. Note that the resulting curves have been scaled to the same magnitude.}
    \label{fig:spin_correlations}
\end{figure}

When applying this method to obtain the spin currents from our simulations on RuO$_2$, we consider only the spin transport to the respective nearest unit cells to simplify the calculations. As demonstrated in Fig.~\ref{fig:spin_correlations}a, the intra-sublattice correlations turn out to be much larger than the inter-sublattice correlations, i.e., $\braket{\vec{S}_{is}\times \vec{S}_{i\pm1,s'=s}} \gg\braket{\vec{S}_{is}\times \vec{S}_{i\pm1,s'\neq s}}$.
This is due to the fact that only in the vicinity of the $\Gamma$ point of the Brillouin zone the magnon excitations involve both sublattices while they are primarily hosted by only one of them at the rest of the Brillouin zone.

Henceforth, we use the following expression for estimating spin currents in RuO2,
\begin{align}
    \label{eq:app_spin_current}
    \vec{j}_i
    &\approx
    -
    \frac{\gamma}{\mu_\mathrm{s}}
    A_\mathrm{lat}
    J
    \Big(
    \Big\langle
    \vec{S}_{i\mathrm{a}}
    \times
    \vec{S}_{i+1,\mathrm{a}}
    \Big\rangle
    +
    \Big\langle
    \vec{S}_{i\mathrm{b}}
    \times
    \vec{S}_{i+1,\mathrm{b}}
    \Big\rangle
    \Big),
\end{align}
where we have introduced $J:=J_{i,i+1}^\mathrm{aa}=J_{i,i+1}^\mathrm{bb}$. We checked in Fig.~\ref{fig:spin_correlations}b that neglecting intersublattice correlations indeed has no qualitative impact on the spin current profile.

%

\end{document}